# MOTION BACKWARDS IN TIME?

# A SECOND EXAMPLE

**Evangelos Chaliasos**

365 Thebes Street

GR-12241 Aegaleo

Athens, Greece

**It is a striking result the one found by Wang, Kuzmich & Dogariu [1] (see also [2]) experimentally last year, that a pulse of light, entered into an atomic caesium (Cs) vapour cell, appeared at the exit *before* its entrance (!) by 62 nanoseconds. It has to be mentioned that the shape of the pulse was maintained (cf. fig. 4 of [1]), so that its peak was not shifted forward inside the pulse. Thus the explanation that the peak of the pulse was advanced at the exit, with the result of emerging there before the entrance of the peak of the incident pulse, does not apply.**

This "strange" effect was due to an observed negative group-velocity index $n_g \approx$ $\approx -330$ (while the group-velocity index implied by the 62 nanoseconds is $n_g \approx -310$). In order to achieve this index the researchers mentioned used a double emission line, namely the 852 nm caesium $D_2$ line, and found how the optical refractive index $n(\nu)$ had to vary between (and closed to) the two lines of the doublet theoretically, constructing a theoretical curve which was calculated from the appropriate theoretical formula (eq. (1) of [1]) after giving to the parameters *values obtained experimentally.*[*] This curve fitted perfectly the experimental data on values of the optical refractive index in the anomalous dispersion region studied. But because of the encountered sufficiently negative slope of n(ν), almost constant, between the two lines of the doublet, a negative $n_g$ resulted ($n_g \approx -330$) from the formula giving it, namely $n_g = n(\nu) + \nu dn(\nu)/d\nu$. And it is evident that a negative group-velocity index cannot be derived theoretically by any known means, starting from causality, as we will see in what follows.

---

[*] For the full theory applicable to such an anomalous dispersion region see [3] (§ 15-5) (for example).

It is argued by Wang, Kuzmich & Dogariu that this "strange" effect can be explained by the wave nature of light, without having to violate causality. They write[1]: "…interference between different frequency components [of the pulse] produces this rather counterintuitive effect". They clearly tacitly assume that the information travels at the phase-velocity speed. But this is not correct. It is evident that information, like energy, travels at the speed of the pulse as a whole (evenmore since its shape does not change). And the latter is nothing else than the velocity of propagation of the pulse, which has to be taken equal to its *group-velocity*[4](p.291), which can never become negative for motion forward in time, as we will see in what follows.

Also, it has to be noted that the equation describing the anomalous dispersion region (eq. (1) of [1]) is indeed resulting from the Kramers-Kronig relations[4,5], which in turn is based undoubtedly on causality, described by the relation $ct - x > 0$, whence $0 < x/t < c$, where x is the distance traveled by the pulse in time t ($x > 0$). This is to say that the pulse´s front travels with positive velocity less than c. This pulse´s velocity is just the *group-velocity* $d\omega/dk = x/t$, where ω is the cyclic frequency and k is the wave number. Thus, because of causality, we have to impose the condition $0 < \upsilon_g \equiv d\omega/dk < c$, for $x > 0$. In other words, causality establishes that the signal cannot propagate with a velocity greater than c (cf. [5] §7.11c). Of course we could have $\upsilon_g$ negative, this fact meaning that the pulse propagates from the receiver to the emitter

( x < 0 ). But, because the emitter looses energy while the receiver gains energy, this is impossible. Thus, in the case of negative $\upsilon_g$ here, the only thing that we can conclude is that the pulse propagates from the emitter to the receiver ( x > 0 ) but *moving backwards in time!* ( t < 0 ). Of course, on the other hand, for the group-velocity index $n_g$, given by $n_g = c / \upsilon_g$, we should have $n_g > 1$ ( > 0 ). In other words, for x > 0 , we can *never* have $\upsilon_g < 0$ (or $n_g < 0$ ), if we insist on causality!

Thus the condition ct – x > 0 (with x > 0) (causality) must necessarily lead to positive $\upsilon_g$ (and $n_g$), that is to x and t of the same sign (so t > 0). This means that the very fact of Wang, Kuzmich & Dogariu finding a negative $\upsilon_g$ (and $n_g$) (with x > 0) undoubtedly implies a *violation of causality* in the particular case! The big problem is then to understand the evidently non-causal character of the effect discovered experimentally by Wang, Kuzmich & Dogariu. And to this end, an insight is given by the invariance of the space-time interval (in one dimension), namely by the fact that the expression $c^2 t^2 – x^2$ retains the same value independently of the frame used [6] ( § 2 ). Thus, in order for causality to hold, it is necessary that $c^2 t^2 – x^2 > 0$, that is the space-time interval to be *time-like*. But this relation can be written as ( c t – x) · · ( c t + x) > 0 . Thus, it is necessary that the binomials c t – x  and  c t + x  are of the same sign. The case where c t – x > 0  &  c t + x > 0  right describes what we call *causality*. If $\upsilon$ is the velocity, then, for x > 0 , c t – x > 0 implies $\upsilon < 0$ and, for x < 0, c t + x > 0  implies  $\upsilon > - c$ , that is causality implies  $- c < \upsilon < c$ . For motion in the positive x-direction of course $\upsilon > 0$ , while for motion in the negative x-direction we will have $\upsilon < 0$ .

But for the space-time interval $c^2 t^2 – x^2$ to be time-like ( > 0) it is not necessary for both c t – x  &  c t + x to be positive (causality). We can evidently have also c t – - x < 0  &  c t + x < 0 , from a theoretical point of view. We may then say that, in this case, a principle of *anti-causality* holds (for particles obeying these two conditions). If then again $v$ is the velocity of such a particle, for $x < 0$, the condition $ct - x < 0$ (or $-ct + x > 0$) implies that $v < c$, while, for $x > 0$, the condition $ct + x < 0$ (or $-ct - x > 0$) implies that $v > -c$, that is again anti-causality (now) implies $-c < v < c$. But now for motion in the positive x-direction $v < 0$, while for motion in the negative x-direction $v > 0$. This is the case because now $t < 0$, that is we have motion *backwards in time!*

Combination of both principles of causality and of anti-causality would give a principle of *generalized causality* as $(ct - x)(ct + x) > 0$, or $c^2 t^2 - x^2 > 0$, which would state simply that the radius vector $(ct, x)$ be *time-like!* Then of course both principles, once they were valid in one frame, they would be valid in any other frame, because of relativistic covariance. Thus, *tachyons* would not follow either of the two principles, that is the generalized one, as defining space-like radius vectors.

In another way, according to the generalized causality principle, namely $c^2 t^2 - x^2 > 0$, we would simply have $c > |x/t|$, or again $-c < v < c$ for any allowed velocity. And if we have $v > 0$ (with x positive) we will have to do with motion forward in time ($t > 0$), while, if we had $v < 0$ (with x positive), we would have to do with motion *backwards in time!* ($t < 0$). The latter is clearly the case on hand, since we have found $v < 0$ (x & t of opposite signs, with x positive). The specific value found, namely $v = -c/310$, clearly satisfies the condition $-c < v < 0$ ($x > 0$), that is the principle of *anti-causality* holds in our case! Concerning the refractive index, since $-c < v_g < c$, it is easy to see that $n_g > 1$ ($x > 0$) for motion forward in time, while $n_g < -1$ ($x > 0$) for motion backwards in time. The latter condition is satisfied here, since $n_g = -310$.

A last remark only. In the case of motion backwards in time (anti-causality), it can be easily shown that the Kramers-Kronig relations become:

$$\text{Re } n(\omega) + 1 = (2/\pi) P \int d\acute{\omega} \, [\acute{\omega} \text{ Im } n(\acute{\omega})] / [\acute{\omega}^2 - \omega^2]$$

$$\text{Im } n(\omega) = -(2\omega/\pi) P \int d\acute{\omega} \, [\text{Re } n(\acute{\omega}) + 1] / [\acute{\omega}^2 - \omega^2],$$

where P means "principal part", and the integrations are taken from 0 to $-\infty$. I intend to prove them in another paper.